\documentclass[11pt]{article}
\usepackage{caption}
\usepackage{fa2025}
\usepackage{amsmath}
\usepackage{cite}
\usepackage{url}
\usepackage{subcaption}
\usepackage{graphicx}
\usepackage{color}
\usepackage{siunitx}
\usepackage[utf8]{inputenc}
\usepackage{booktabs}
\usepackage{array}
\usepackage{tabularx}

\title{Psychoacoustic assessment of synthetic sounds for electric vehicles in a virtual reality experiment}

\multauthor
{Pavlo Bazilinskyy,$^{1*}$ \hspace{1cm}Md Shadab Alam$^1$ \hspace{1cm} Roberto Merino-Martínez$^2$} { \\
$^1$ Department of Industrial Design, Eindhoven University of Technology, The Netherlands\\
$^2$ Faculty of Aerospace, Delft University of Technology, The Netherlands\\
\correspondingauthor{p.bazilinskyy@tue.nl}{Bazilinskyy et al.}
}

\begin{document}

\maketitle
\begin{abstract}
The growing adoption of electric vehicles, known for their quieter operation compared to internal combustion engine vehicles, raises concerns about their detectability, particularly for vulnerable road users. To address this, regulations mandate the inclusion of exterior sound signals for electric vehicles, specifying minimum sound pressure levels at low speeds. These synthetic exterior sounds are often used in noisy urban environments, creating the challenge of enhancing detectability without introducing excessive noise annoyance. This study investigates the design of synthetic exterior sound signals that balance high noticeability with low annoyance. An audiovisual experiment with 14 participants was conducted using 15 virtual reality scenarios featuring a passing car. The scenarios included various sound signals, such as pure, intermittent, and complex tones at different frequencies. Two baseline cases, a diesel engine and only tyre noise, were also tested. Participants rated sounds for annoyance, noticeability, and informativeness using 11-point ICBEN scales. The findings highlight how psychoacoustic sound quality metrics predict annoyance ratings better than conventional sound metrics, providing insight into optimising sound design for electric vehicles. By improving pedestrian safety while minimising noise pollution, this research supports the development of effective and user-friendly exterior sound standards for electric vehicles.
\end{abstract}
\section{Introduction}
\label{sec:introduction}
The World Health Organisation reports that pedestrians account for 21\% of the 1.19 million annual traffic deaths (roughly 250,000) \cite{who2023}. A significant proportion of these casualties occur during road crossings \cite{schneider2016}. Factors contributing to such incidents include misjudging the time required to cross, low visibility conditions, and visual obstructions that prevent the timely detection of approaching vehicles \cite{oxley2005, rosenbloom2004, dommes2013}. Enhancing auditory cues emitted by vehicles has been proposed as a potential strategy to mitigate these risks, particularly in the context of typically quieter electric vehicles (EVs) \cite{bazilinskyy2023exterior} and external Human-Machine Interfaces (eHMIs) for automated vehicles \cite{cabrall2020,Deb2024,Rodriguez2024,Sabic2019}.

The primary purpose of adding synthetic sounds to EVs is to enhance their noticeability, particularly for vulnerable road users (VRUs), such as pedestrians (and especially individuals with visual impairments and hearing problems). Lai \textit{et al.} indicated that sound level, temporal characteristics, and modulation patterns affect the speed with which pedestrians can detect an approaching vehicle \cite{Lai2023}. Effective sound design should maximise noticeability at low speeds without being overbearing in quieter environments. Psychoacoustic metrics such as loudness, sharpness, roughness, and fluctuation strength provide valuable information to optimise detectability while maintaining perceived annoyance within acceptable limits \cite{Fiebig2020}.

Annoyance is a critical factor in designing synthetic sounds for EVs, as excessive noise levels can cause discomfort and increased environmental noise pollution. Studies have shown that the spectral content, duration, and repetition of sound signals influence perceived annoyance \cite{Bodden2016, Lai2023,Osses2016}. Intermittent sounds, for example, can be perceived as more intrusive than continuous ones \cite{Gwak2014}, whereas high-frequency sounds tend to be more irritating than lower-frequency ones \cite{vonBismark1974}. Therefore, optimising EV sounds requires a balance between noticeability and minimal disturbance to urban residents \cite{Burdzik2022, Heinrichs2003}.

An ideal EV sound should convey useful information to VRUs on vehicle behaviour, such as acceleration or deceleration. Informative sounds help users make more accurate judgments about vehicle proximity and direction of movement \cite{Fiebig2020}. Complex sound structures, which integrate variations in pitch and modulation, can enhance informativeness while maintaining societal acceptance. Designing EV sounds that intuitively communicate vehicle dynamics can improve pedestrian interaction with EVs in urban settings \cite{Lai2023}.

This preliminary study aims to examine the psychoacoustic noise annoyance of various sounds of exterior vehicles to inform pedestrians of an approaching vehicle. In a virtual reality (VR) experiment, participants were asked to rate this parameter using the 11-point ICBEN scale. The open-source MATLAB toolbox SQAT (Sound Quality Analysis Toolbox) v1.2 \cite{SQAT_Zenodo_v1.2} toolbox was then employed to calculate different psychoacoustic metrics to predict annoyance ratings. This analysis provides useful insights into optimising the design of external auditory communication for EVs.

Section~\ref{sec:method} discusses the methodology employed, including the synthetic sounds considered, the VR experiment, and the sound metrics used. The main results are presented and discussed in section~\ref{sec:results}, whereas the conclusions and recommendations for future work are gathered in Section~\ref{sec:conclusions}. Finally, supplementary material is provided in Section~\ref{sec:supp_mat}.


\section{Methodology}
\label{sec:method}

\subsection{Synthetic sounds}
Table~\ref{tab:sound_characteristics} lists the 15 synthetic sounds that were used in the study, which can be categorised into four groups: (1) continuous pure tones at a single frequency, (2) intermittent pure tones with alternating 500-ms on/off intervals, (3) combined tones at a principal frequency plus secondary tones at frequencies $\pm$90~Hz from the principal tone, and (4) double beeps at 1800–1900~Hz. The double beeps consisted of a 240~ms beep, a 10~ms pause, a 240~ms beep, and a 1000~ms pause. Furthermore, a diesel engine sound\footnote{\url{https://youtu.be/watch?v=2Y33bTlAA-E}} was included as a reference case for internal combustion engine vehicles due to its distinctive and more familiar noise. To evaluate the sound profile of a quiet electric or automated vehicle, a stimulus containing only tyre noise\footnote{\url{https://youtu.be/watch?v=X0wpizkkH_Q}} (i.e., without any synthetic sound added) was also incorporated. All evaluated sound stimuli were combined with a background noise recording of a quiet street\footnote{\url{https://youtu.be/watch?v=6C-W_7BZBxQ}}. This list of sounds was previously used in an online crowdsourcing listening experiment (i.e. without VR) reported in \cite{bazilinskyy2023exterior}.

\begin{figure} 
 	\centering
	\includegraphics[width=.8\linewidth]{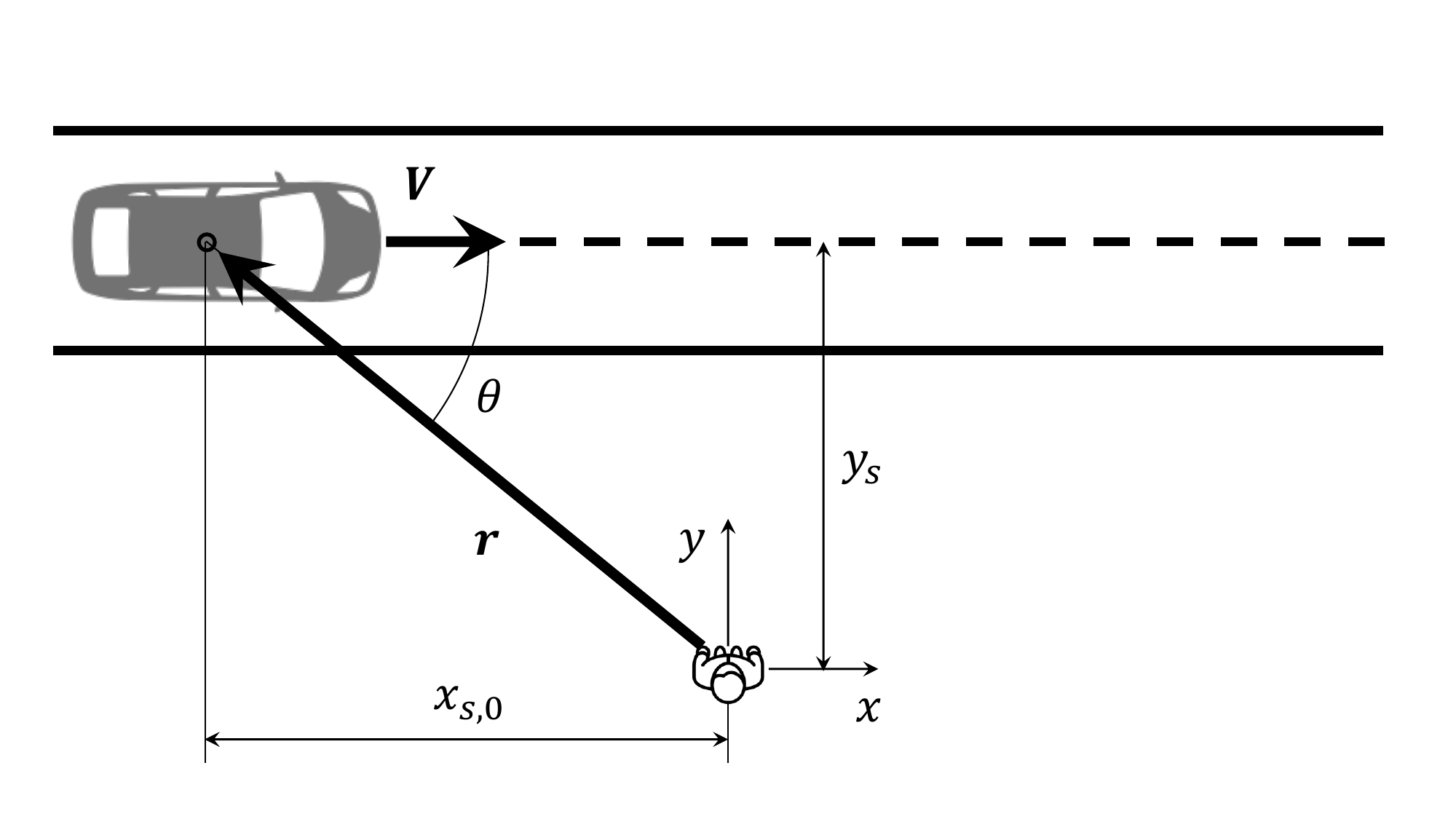}
	\caption{\label{fig:diagram-vehicle}Scheme of the considered sound source and observer geometry \cite{bazilinskyy2023exterior}.}
\end{figure}

This study considers a relatively simple sound source and observer geometry in a two-dimensional arrangement, i.e., the observer and the sound source are on the same plane; see Fig.~\ref{fig:diagram-vehicle}. For simplicity, the sound source is considered a point source with a constant velocity of 30~km/h (i.e., 8.33~m/s) in the positive $x$ direction $\bm{V}=(V_{x},\, 0)$, i.e., the source has a linear trajectory along a straight line at a distance $y_{s}$ of 3~m from the observer, see Fig.~\ref{fig:diagram-vehicle}. The initial position of the sound source at is defined at $\bm{r} (0)=(x_{s,0}, \, y_{s})$ and at a general instant $\tau_{e}$ as $\bm{r} (\tau_{e})=(x_{s,0} + V_{x}\tau_{e}, \, y_{s})$. For the purpose of sound signal generation, the source moved from $x_{s,0}=-60$~m to $x_{s}=60$~m. Therefore, a signal length of 14.4 s was generated and used in each case. All sound stimuli (except for the case with only tyre noise) had an equivalent A-weighted sound pressure level ($\text{L}_{\text{p,A,eq}}$) of 65 dBA in the observer position. The VR environment employed enables binaural rendering of the sound accounting for the relative position of the sound source, the orientation of the participant's head, and the Doppler-shift due to relative motion. The reflection of sound on the ground or any reflecting surfaces (such as buildings) was neglected for simplicity. 

\begin{table}[!h]
\caption{Synthetic sounds used in the experiment.}
\vspace{4mm}
\begin{center}
\begin{tabular}{|c|p{6.9cm}|}
\hline
\textbf{No.} & \textbf{Characteristics} \\ \hline
1  & Pure tone, continuous, 350~Hz \\ \hline
2  & Pure tone, continuous, 500~Hz \\ \hline
3  & Pure tone, continuous, 1000~Hz \\ \hline
4  & Pure tone, continuous, 2000~Hz \\ \hline
5  & Pure tone, intermittent (13 $\times$ [500~ms emitting, 500~ms not emitting]), 350~Hz \\ \hline
6  & Pure tone, intermittent (13 $\times$ [500~ms emitting, 500~ms not emitting]), 500~Hz \\ \hline
7  & Pure tone, intermittent (13 $\times$ [500~ms emitting, 500~ms not emitting]), 1000~Hz \\ \hline
8  & Pure tone, intermittent (13 $\times$ [500~ms emitting, 500~ms not emitting]), 2000~Hz \\ \hline
9  & Combined tone, continuous, 350~Hz ($\pm$90~Hz) \\ \hline
10 & Combined tone, continuous, 500~Hz ($\pm$90~Hz) \\ \hline
11 & Combined tone, continuous, 1000~Hz ($\pm$90~Hz) \\ \hline
12 & Combined tone, continuous, 2000~Hz ($\pm$90~Hz) \\ \hline
13 & Double beeps (8 $\times$ [240~ms beep, 10~ms pause, 240~ms beep, 1000~ms pause]), 1800–1900~Hz \\ \hline
14 & Diesel engine \\ \hline
15 & Tyres on asphalt \\ \hline
\end{tabular}
\end{center}
\label{tab:sound_characteristics}
\end{table}

\begin{figure}[ht]
 \centerline{
 \includegraphics[width=8cm]{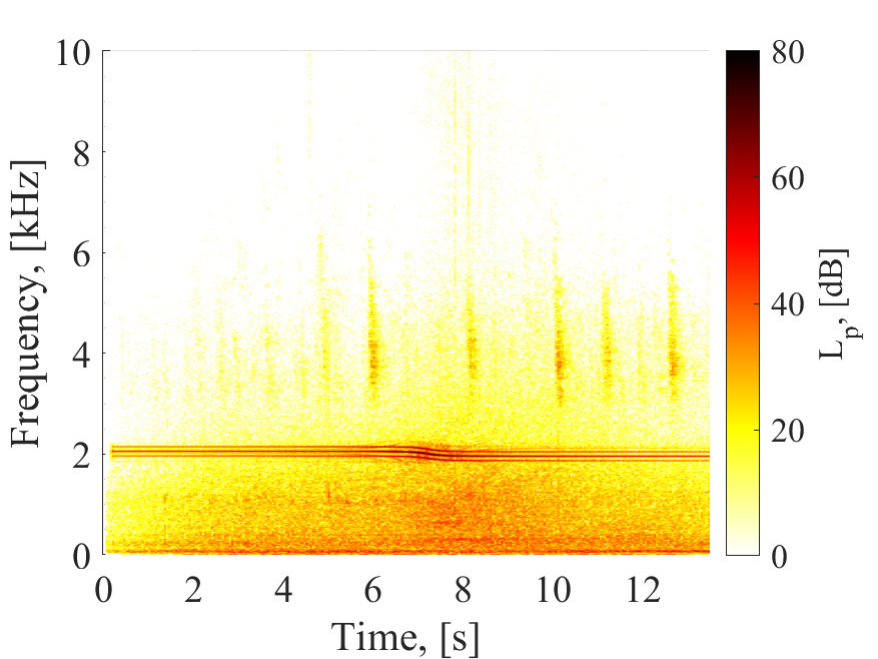}}
 \caption{Spectrogram for the sound stimuli Combined tone, continuous, 2000~Hz (case No. 12).}
 \label{fig:spectrogram_comb_tone2000Hz}
\end{figure}

An exemplary spectrogram is provided in Fig.~\ref{fig:spectrogram_comb_tone2000Hz} for test case No. 12 (Combined tone, continuous, 2000~Hz), where the tone at said frequency (plus the two secondary tones at $\pm$90~Hz) is clearly identified as the primary sound source. In addition, broadband noise is visible, especially below 3~kHz, mostly due to the simulated noise of the tyres on the asphalt and background noise. The motion of the source can also be observed in Fig.~\ref{fig:spectrogram_comb_tone2000Hz} in the slight Doppler frequency shift and in the higher levels observed when the source is closer to the observer (roughly around half the recording time). The short vertical lines centred around 4~kHz between 6~s and 13~s correspond to birds tweeting within the background noise recording.

\subsection{Virtual reality experiment}

\begin{figure}[ht]
  \centering
  \includegraphics[width=7cm]{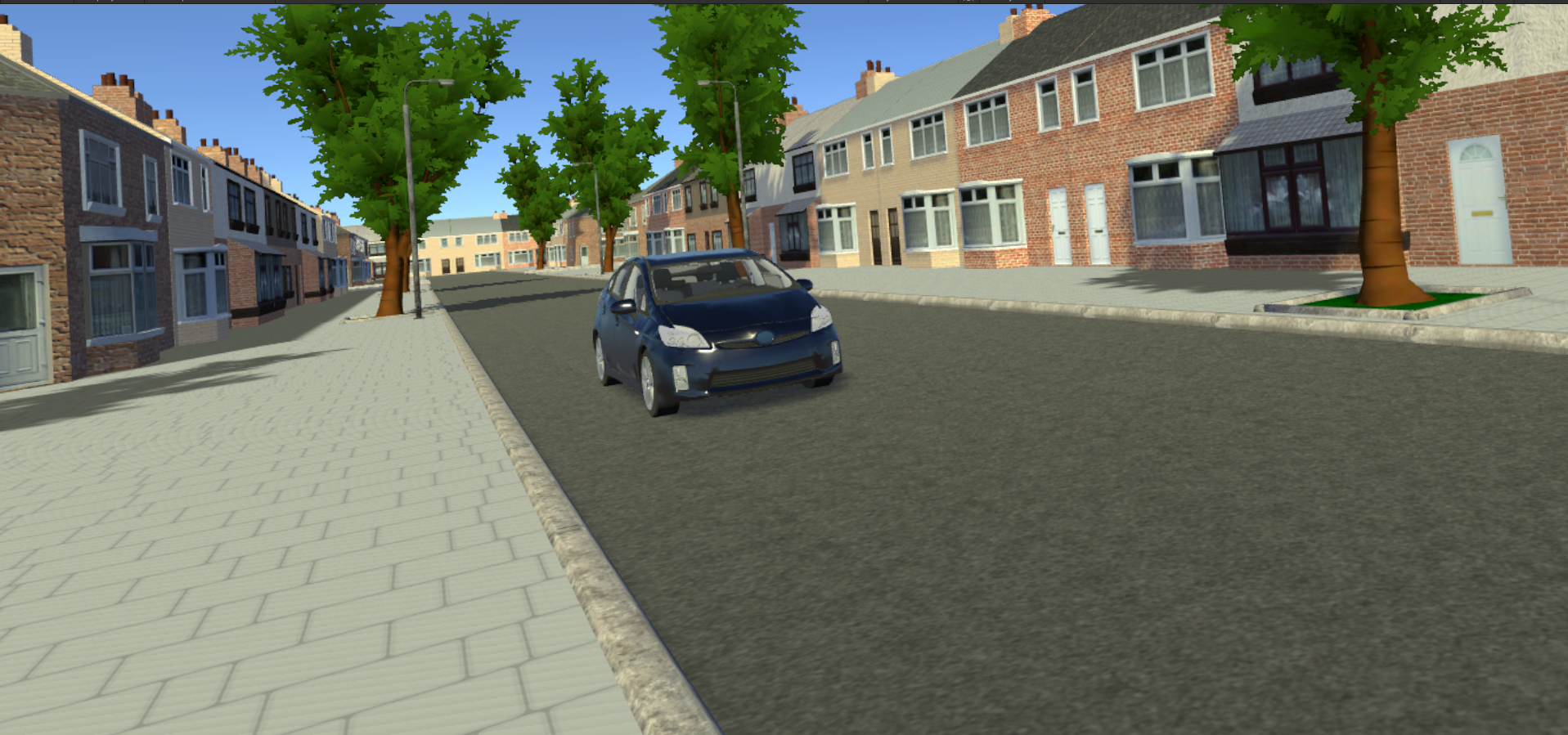}
  \hspace{0.5cm}
  \includegraphics[width=7cm]{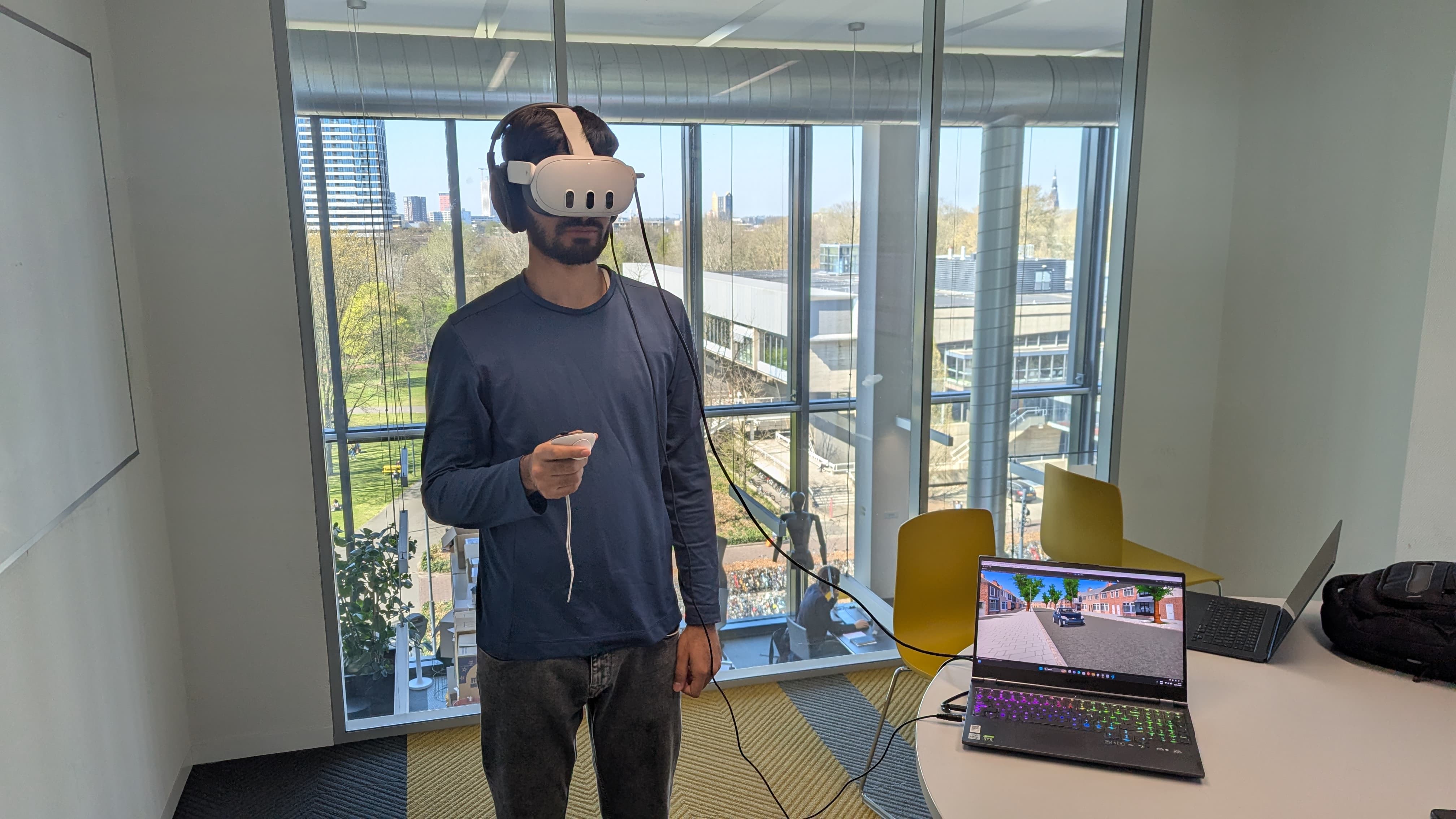}
  \caption{(Top) Example view of participant in VR environment. (Bottom) VR experiment at Eindhoven University of Technology.}
  \label{fig:vr-setup}
\end{figure}

\begin{figure}[ht]
 \centerline{
 \includegraphics[width=7.2cm]{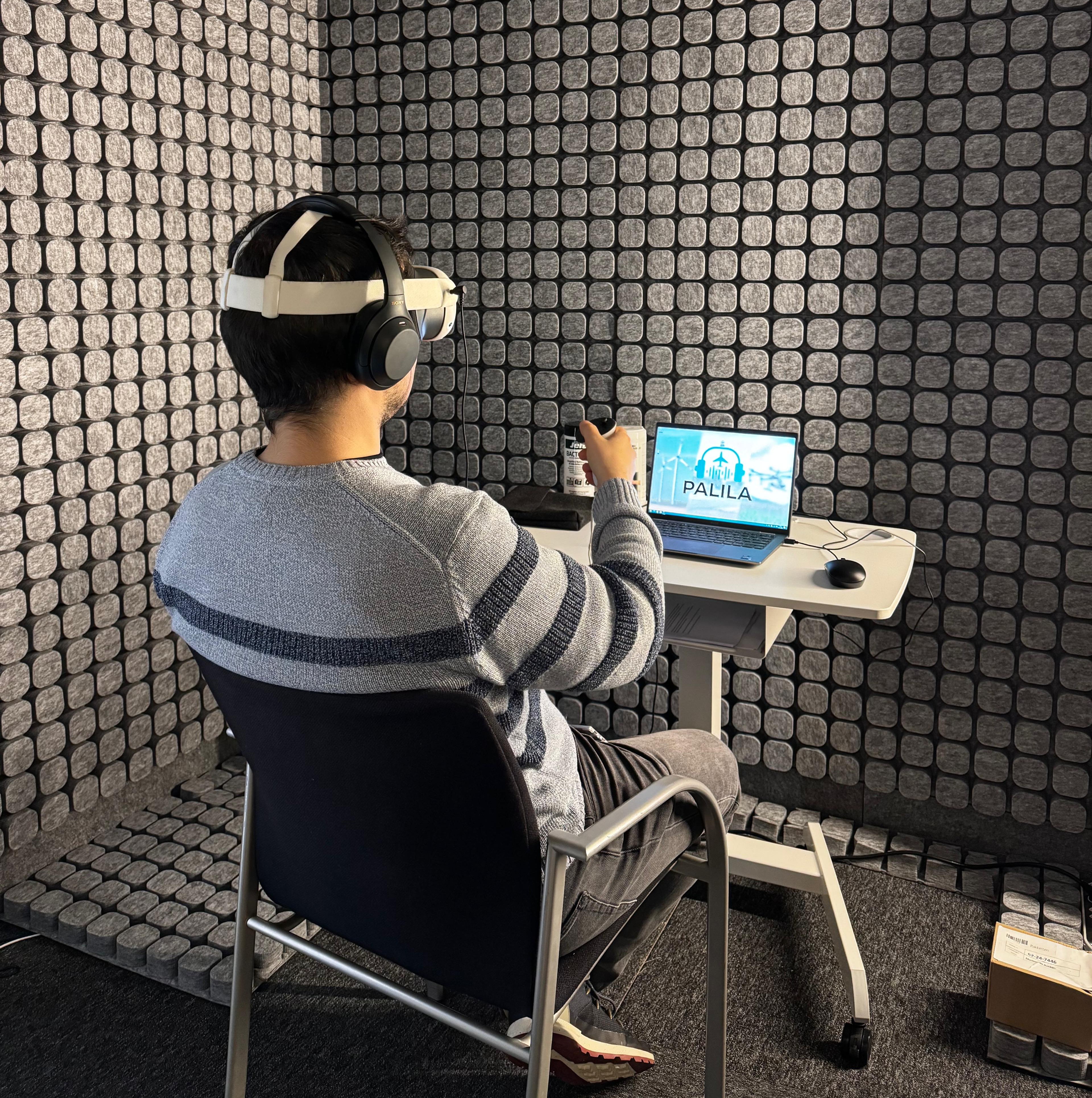}}
 \caption{Example of a VR experiment in PALILA at Delft University of Technology. In practice, participants stood up during the experiment.}
 \label{fig:vr_PALILA}
\end{figure}

A total of 14 participants (11 males and 3 females) with an average age of 30 years (SD = 4.63 years) from Eindhoven University of Technology and Delft University of Technology participated in the experiment. In both universities, the experiment was conducted in a quiet environment and with the same equipment. In particular, the sessions at Delft University of Technology were performed in the Psychoacoustic Listening Laboratory (PALILA) \cite{Merino-Martinez2023}, see Fig.~\ref{fig:vr_PALILA}. The study was approved by the Ethics Review Boards of both universities, and the participants gave their informed consent to use their data.

The VR setup was developed using Unity 2022.3.5f1 (see Section~\ref{sec:supp_mat}). Participants wore a Meta Quest 3 head-mounted display (HMD), which was connected via cable to a Lenovo Legion 81YT laptop featuring an Intel Core i7-10750H 2.60GHz CPU, 32.0 GB RAM and an NVIDIA GeForce RTX 2080 Super graphics card. The audio stimuli were reproduced using Sennheiser HD 560S headphones. Figure~\ref{fig:vr-setup} shows the view of the participant. The ambient noise (measured as $\text{L}_{\text{p,A,eq}}$) during the sessions in Eindhoven was 35.7 dBA, while in PALILA it was 13.4 dBA.

At the beginning of the experiment, participants completed a consent form and provided demographic details such as gender, age, and nationality. Subsequently, the HMD was reset to clear boundary configurations and a spatial scan history. The floor level was calibrated by the experimenter using the HMD, after which the participants were then asked to stand at a predefined location within the laboratories. The participants were then handed the HMD to establish their virtual boundary, ensuring an accurate recording of the participant's height and spatial location within the VR system. The HMD was then connected to the laptop via Quest Link. The Unity application was launched, and the experiment began after activating the play button within Unity.

The participants received instructions at the beginning of the experiment inside the VR, which were as follows: ``\textit{Imagine that you are a pedestrian standing on the side of the road. You will experience 15 audiovisual scenarios of a vehicle driving by you. During each scenario, press and HOLD the trigger when you feel safe to cross the road in front of the car. You can release the button and then press it again multiple times during the scenario. After each scenario, you will be asked to answer a few questions. Press the button to proceed. The experiment will start with a training scenario to familiarise yourself with the environment. During this scenario, press and HOLD the trigger when you feel safe crossing the road in front of the car. You can release the button and then press it again multiple times during the scenario. Press the button to start.}". The keypress data is not presented in this paper but instead will be employed for future research investigating safety, see Section~\ref{sec:conclusions}.

A preliminary training trial was conducted to allow participants to acclimate to the VR setup and practice the response task. Subsequently, the participants completed 15 experimental trials presented in random order to mitigate learning effects. During these trials, participants were asked to indicate their willingness to cross the road in front of the vehicle by following the instruction: ``\textit{Start by HOLDING the trigger button. Release the trigger button when it becomes unsafe to cross; press it again when safe to cross}". Following each trial, including the test scenario, the participants were presented with three questions using 11-point slider scales (from 0 to 10) displayed within the VR environment. The question assessed noticeability (``\textit{The vehicle sound was easy to notice (0 = not easy to notice, 10 = easy to notice)}"), informativeness (``\textit{The sound gave me enough information to realise that a vehicle was approaching (0 = not enough information, 10 = enough information)}") and annoyance (``\textit{The vehicle sound was annoying (0 = not annoying, 10 = extremely annoying)}"). For the present research, only the results of the last question (related to annoyance) are considered.

Upon completion of all the trials, the HMD was removed and the participants completed a post-experiment questionnaire (see Section~\ref{sec:supp_mat}). Each experimental session lasted approximately 15 minutes on average. Following the session, the participants received a €10 voucher as compensation for their time and participation.


\subsection{Conventional and sound quality metrics}
\label{sec:SQM}
Conventional sound metrics typically used in noise assessment pose challenges in quantifying noise annoyance \cite{Merino-Martinez2024d,Merino-Martinez2022c}. However, current noise regulations still employ these metrics to enforce environmental noise laws. Therefore, the current study considers the maximum sound pressure level $L_{\text{p,max}}$, as well as its A-weighted version, $L_{\text{p,A,max}}$. In addition, the maximum tone-corrected perceived noise level ($\text{PNLT}_{\text{max}}$) and the effective perceived noise level (EPNL), which are typically used in evaluating aircraft noise \cite{Merino-Martinez2025}, were included in the analysis.

Unlike the sound pressure level $L_{\text{p}}$ metric, which quantifies the purely physical magnitude of sound based on pressure fluctuations, sound quality metrics (SQMs) describe subjective perception of sound by human hearing. Hence, SQMs are expected to better capture the auditory behaviour of the human ear compared to conventional sound metrics typically employed in noise assessments \cite{Merino-Martinez2025}. The five most commonly used SQMs \cite{Greco2023} are:

\begin{itemize}
    \item Loudness (N): Perception of the magnitude of the sound corresponding to the overall intensity of the sound. 
    \item Tonality (K): Perceived strength of unmasked tonal energy within a complex sound. 
    \item Sharpness (S): High-frequency sound content. 
    \item Roughness (R): Hearing sensation caused by modulation frequencies between 15~Hz and 300~Hz. 
    \item Fluctuation strength (FS): Assessment of slow fluctuations in loudness with modulation frequencies up to 20~Hz, with maximum sensitivity for modulation frequencies around 4~Hz. 
\end{itemize}

These five SQMs were calculated for each sound wave and combined into a single global psychoacoustic annoyance (PA) metric following the model proposed by Di \textit{et al.} \cite{Di2016}. Henceforth, the top 5\% percentiles of these metrics (values exceeded 5\% of the time) are reported (and hence the subindex \textit{5}). All SQMs and the PA metric were computed using the open-source MATLAB toolbox SQAT (Sound Quality Analysis Toolbox) v1.2 \cite{SQAT_Zenodo_v1.2}.


\section{Results and discussion}
\label{sec:results}
\begin{figure}[ht]
 \centerline{
 \includegraphics[width=8cm]{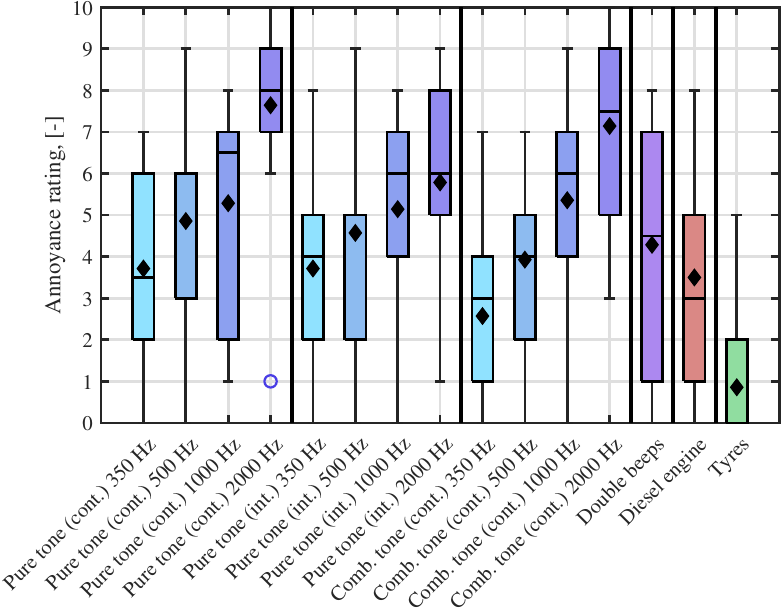}}
 \caption{Annoyance ratings per sound stimulus.}
 \label{fig:box_plot_annoyance}
\end{figure}

The annoyance ratings per scene are shown in Fig.~\ref{fig:box_plot_annoyance} as a box plot. In each box, the diamond marker denotes the mean value, the central horizontal line denotes the median values, the edges of the box are the $25^{\text{th}}$, and the $75^{\text{th}}$ percentiles (also known as the interquartile range, IQR), and the whiskers extend to the most extreme data points. Outliers are plotted individually as circles. The different groups of test cases (pure tones, combined tones, and intermittent tones) are separated with vertical black lines and ordered by increasing tonal frequency.

In general, all tonal cases (pure tones, intermittent tones and combined tones) show the same consistent trend of increasing perceived annoyance as the tonal frequency increases (from about 3/10 on average for the 350~Hz cases to roughly 7/10 ratings for the 2~kHz cases). This is somewhat expected due to the influence of the increased sharpness of the stimuli on the annoyance \cite{vonBismark1974}. On average, pure tones were perceived as more annoying (5.4 mean rating) than intermittent and combined tones, as both cases have mean ratings of 4.8/10. The double beep case presents a particularly large spread in annoyance ratings, with an IQR ranging from 1 to 7 and both median and mean values around 4.5. The diesel engine stimulus shows, in general, lower mean and median annoyance ratings (around 3.5 and 3, respectively) than most tonal cases. The higher familiarity with this type of sound might have caused the relatively lower perceived annoyance. Lastly, as expected, the case that only featured tyre noise was perceived as the least annoying (with mean annoyance ratings around 1).

\begin{figure}[ht]
 \centerline{
 \includegraphics[width=8cm]{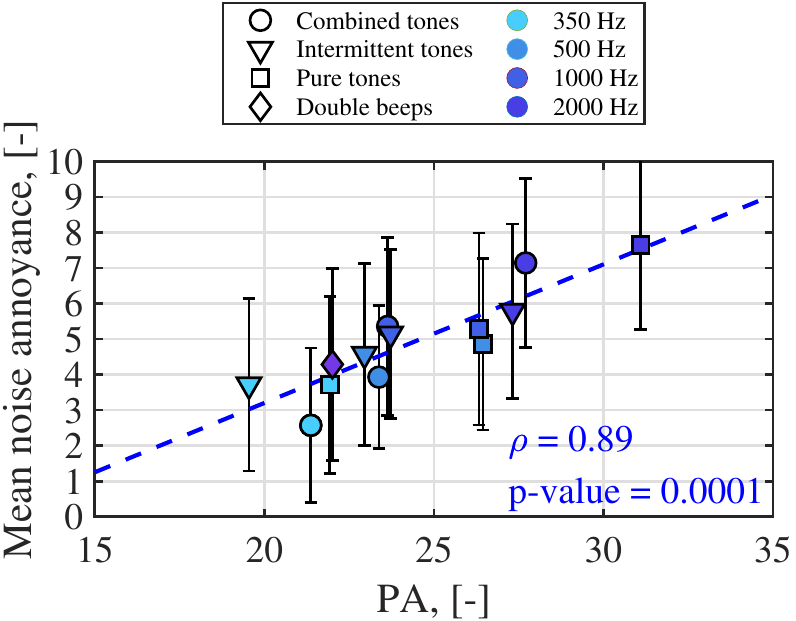}}
 \caption{Correlation analysis between the mean annoyance ratings and the psychoacoustic annoyance (PA) metric from the model by Di \textit{et al.} \cite{Di2016}. The error bars represent the standard deviation of the ratings. The dashed blue line denotes the least-squares fit.}
 \label{fig:annoyance_vs_PA}
\end{figure}

Table \ref{tab:correlation_table} presents the results of a correlation analysis between the mean annoyance ratings reported in the VR experiment and the different sound metrics, such as Pearson's correlation coefficients $\rho$ and the corresponding p-values. For this analysis, the two signals that corresponded to the diesel engine and the tyres were excluded because they did not represent a synthetic sound signal for EVs and, since they were outliers, negatively influenced the findings. Correlations with p-values greater than 0.05 are considered non-statistically representative \cite{Merino-Martinez2018}. With this criterion, the metrics $L_{\text{p,max}}$, $\text{K}_{\text{5}}$, $\text{R}_{\text{5}}$, and $\text{FS}_{\text{5}}$ are deemed not representative (at least individually) of the variance in the reported annoyance ratings. On the other hand, $L_{\text{p,A,max}}$ presents a strong and significant correlation $\rho \approx 0.8$, similar to the metric $\text{PNLT}_{\text{max}}$. The EPNL and $\text{N}_{\text{5}}$ metrics show slightly higher and comparable predictive performance with $\rho$ values of 0.86 and 0.88, respectively. Interestingly, sharpness $\text{S}_{\text{5}}$ alone also has a strong correlation ($\rho \approx 0.77$) with annoyance ratings. This coincides with the trends observed in Fig.~\ref{fig:box_plot_annoyance}, where all tonal cases presented higher annoyance ratings with increasing tonal frequency. Lastly, the best-performing metric for representing the annoyance rating of the VR experiment was reported to be the PA metric of the model by Di \textit{et al.} \cite{Di2016}, with a value of $\rho - 0.89$, that is, marginally better than those shown by EPNL and $\text{N}_{\text{5}}$. Figure~\ref{fig:annoyance_vs_PA} presents the correlation analysis between this metric and the ratings, together with a least-squares fit with a slope around 0.39~PA.

\begin{table}[!h]
\caption{Pearson correlation coefficients $\rho$ and p-values between each sound metric and the mean annoyance ratings. Non-statistically significant cases (i.e., p-value $>$ 0.05) are denoted in red.}
\begin{center}
\vspace{4mm}
\begin{tabular}{|l|c|c|}
\hline
\textbf{Metric}   & \textbf{$\rho$} & \textbf{p-value} \\ \hline
$\text{L}_{\text{p,max}}$ & -0.3796 & \textcolor{red}{0.2008}                       \\ \hline
$\text{L}_{\text{p,A,max}}$ & 0.8005 & 0.0010                       \\ \hline 
$\text{PNLT}_{\text{max}}$ & 0.7966 & 0.0011                       \\ \hline
EPNL & 0.8623 & 0.0001                       \\ \hline
$\text{N}_{\text{5}}$ & 0.8758 & 0.0001                       \\ \hline
$\text{S}_{\text{5}}$ & 0.7685 & 0.0021                       \\ \hline
$\text{K}_{\text{5}}$ & 0.1770 & \textcolor{red}{0.5631}                       \\ \hline
$\text{R}_{\text{5}}$ & 0.2573 & \textcolor{red}{0.3962}                       \\ \hline
$\text{FS}_{\text{5}}$ & 0.0196 & \textcolor{red}{0.9492}                       \\ \hline
\textbf{PA} & \textbf{0.8866} & \textbf{0.0001}                       \\ \hline
\end{tabular}
\end{center}
\label{tab:correlation_table}
\end{table}

Therefore, it seems that perception-based sound metrics like PA, $\text{N}_{\text{5}}$, and (to some extent) EPNL outperform conventional sound metrics like $\text{L}_{\text{p,A,max}}$. Since all audio stimuli from Fig.~\ref{fig:annoyance_vs_PA} were normalised to have the same $\text{L}_{\text{p,A,eq}}$ of 65~dBA, this metric was not considered in the analysis. Since this metric is normally used in sound assessment, this comparison highlights its intrinsic limitations, as sounds with the same exposure level (in terms of $\text{L}_{\text{p,A,eq}}$) were experienced as considerably different, with mean annoyance ratings ranging from about 2.5 (for combined tone at 350~Hz) to almost 8 (for pure tone at 2000~Hz).

\section{Conclusions and future work}
\label{sec:conclusions}
The current study discussed the findings of a VR experiment featuring a moving car that emits different synthetic sound stimuli to raise awareness of its presence among other road users. This paper focused on performing a psychoacoustic assessment of the sound stimuli employed (e.g., different tonal sounds and beeps) based on the noise annoyance ratings reported by the participants in the experiment. 

In general, pure tones with high tonal frequencies were perceived as the most annoying, whereas intermittent and combined tones with lower tonal frequencies were perceived as more pleasant. Sound quality metrics, such as loudness and sharpness, contribute significantly to the overall annoyance experienced by pedestrians. The psychoacoustic annoyance (PA) model by Di \textit{et al.} \cite{Di2016} provided the best performance for predicting the annoyance ratings of the VR experiments. 

In conclusion, the noise annoyance caused by the synthetic sounds emitted by EVs is an issue that must be addressed thoughtfully during the design process to ensure such vehicles' success and social acceptance. Although synthetic sounds are necessary for safety reasons, their design should carefully balance acoustic comfort with detectability. The results of this investigation seem to indicate that psychoacoustic parameters can predict user perception in a better way than conventional sound metrics that are normally used for noise assessment.

This preliminary study will soon be complemented by analyses of the noticeability and informativeness of the synthetic sounds presented here. To assess the safety per scenario, the willingness of pedestrians to cross in front of an EV emitting synthetic sounds will also be evaluated using the keypress of the VR controller. 

Since this paper focused on rather simple and canonical synthetic sounds, future work may include actual sounds emitted by modern EVs. A larger pool of participants is also desired, ideally including those with visual and/or hearing impairments. Lastly, to raise the applicability of the results to the real world, the VR setup may be replaced by higher-fidelity experimental setups, e.g., an on-road study featuring an EV instrumented with (directional) loudspeakers. 

\section{Supplementary material}
\label{sec:supp_mat}
The VR environment, sounds, materials used in the experiment, analysis code, and anonymised raw data can be found at \url{https://doi.org/10.4121/1f8ae9be-950b-430e-9b75-e2b420dcaa26}. A maintained version of the code is available at \url{https://github.com/Shaadalam9/sound-ev}

\section{Acknowledgments}
\label{sec:acknowledgement}
The research presented in this article is supported and is being carried out within the project \textit{Making sense of sound emitted by electric vehicles} (4TU.NIRICT 2023). Additionally, this publication is also part of the \textit{Listen to the future} project (project number 20247), a part of the Veni 2022 research programme (Domain Applied and Engineering Sciences). The latter project is granted to Roberto Merino-Martinez and is (partially) financed by the Dutch Research Council (NWO).

\bibliography{fa2025_template,PhD_Roberto,bazilinskyy}

%
%
%

\end{document}